\documentclass[10pt,twocolumn,letterpaper]{article}

\usepackage{iccp}
\usepackage{times}
\usepackage{amsmath}
\usepackage{amssymb}
\usepackage{url}
\usepackage{graphicx}
\usepackage{listings}
\usepackage{multirow}
\usepackage{colortbl} 
\usepackage{array} 
\usepackage{hhline}
\usepackage{caption}
\usepackage{subcaption}
\usepackage{dirtytalk}
\usepackage[]{hyperref}


\iccpfinalcopy 

\makeatletter
\newcommand{\printfnsymbol}[1]{%
  \textsuperscript{\@fnsymbol{#1}}%
}
\makeatother

\newcommand{\D}{\mathcal{D}}
\newcommand{\Dtr}{\mathcal{D}_i^{tr}}
\newcommand{\Dts}{\mathcal{D}_i^{ts}}

\def\iccpPaperID{1234} 

\ificcpfinal\fi
\begin{document}

\title{MetaHDR: Model-Agnostic Meta-Learning for HDR Image Reconstruction}

\author{Edwin Pan\thanks{Equal contribution}\\
Stanford University\\
{\tt\small edwinpan@stanford.edu}
\and
Anthony Vento\printfnsymbol{1} \\
Stanford University\\
{\tt\small avento@stanford.edu}
}

\maketitle
\thispagestyle{empty}

\begin{abstract}
Capturing scenes with a high dynamic range is crucial to reproducing images that appear similar to those seen by the human visual system. Despite progress in developing data-driven deep learning approaches for converting low dynamic range images to high dynamic range images, existing approaches are limited by the assumption that all conversions are governed by the same nonlinear mapping. To address this problem, we propose "Model-Agnostic Meta-Learning for HDR Image Reconstruction" (MetaHDR), which applies meta-learning to the LDR-to-HDR conversion problem using existing HDR datasets. Our key novelty is the reinterpretation of LDR-to-HDR conversion scenes as independently sampled tasks from a common LDR-to-HDR conversion task distribution. Naturally, we use a meta-learning framework that learns a set of meta-parameters which capture the common structure consistent across all LDR-to-HDR conversion tasks. Finally, we perform experimentation with MetaHDR to demonstrate its capacity to tackle challenging LDR-to-HDR image conversions. Code and pretrained models are available at \url{https://github.com/edwin-pan/MetaHDR}.
\end{abstract}

\section{Introduction} 

In Low Dynamic Range (LDR) to High Dynamic Range (HDR) conversion, there exists a nonlinear mapping between the radiance in a scene and the recorded pixel values of an image. While existing deep learning models achieve compelling visual results through single-task learning, our insight is that these models implicitly assume that one consistent nonlinear mapping exists for all possible scenes. In reality, every scene has a unique nonlinear mapping dependent on the complex interactions between lighting, camera system, processing, etc. Instead of treating all LDR scene images as part of a single LDR-to-HDR conversion task, our approach treats each LDR scene as a unique task sampled from the task distribution of all possible scenes.

\begin{figure}[htbp!]
    \centering
    \includegraphics[width=0.85\linewidth]{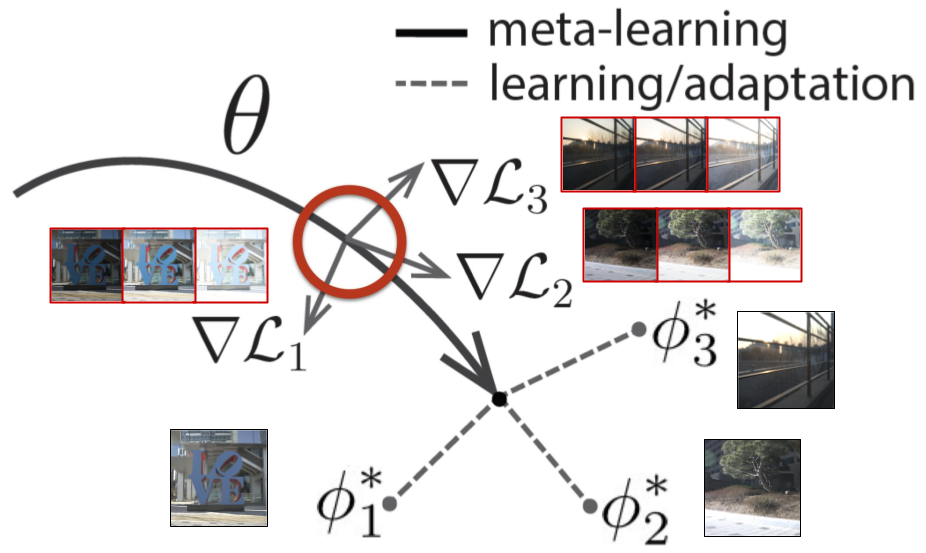}
    \caption{Diagram outlining how MAML optimizes for a general set of meta-parameters $\theta$ in the context of our application.}
    \label{fig:meta-figure}
\end{figure}

To learn the common pattern inherent in all LDR-to-HDR conversion tasks, we propose the use of a meta-learning framework for computing a set of meta-parameters that can quickly be optimized for a specific task given given task-specific information. The MetaHDR model uses model-agnostic meta-learning (MAML)  \cite{finn2017modelagnostic} as the meta-learning framework of choice (see Fig. \ref{fig:meta-figure}).

Through meta-learning, we learn a set of meta-parameters that capture the common patterns between separate nonlinear mappings. This way, given a few specific examples of LDR images, we can quickly adjust the nonlinear mapping to accommodate the specific LDR-to-HDR conversion task at hand.

\section{Related Work}
Conversion from LDR-to-HDR has been a long-standing task. Debevec and Malik developed a method to fuse together multiple LDR images captured at different exposures to create an HDR image \cite{debevec_97}. In their method, each pixel in each LDR image is given a weight depending on how illuminated or saturated it is. All of the LDR images are combined to create one HDR image which will effectively highlight the well-exposed regions of each LDR image.

\begin{figure*}[t!] 
    \centering
    \includegraphics[width=0.85\textwidth]{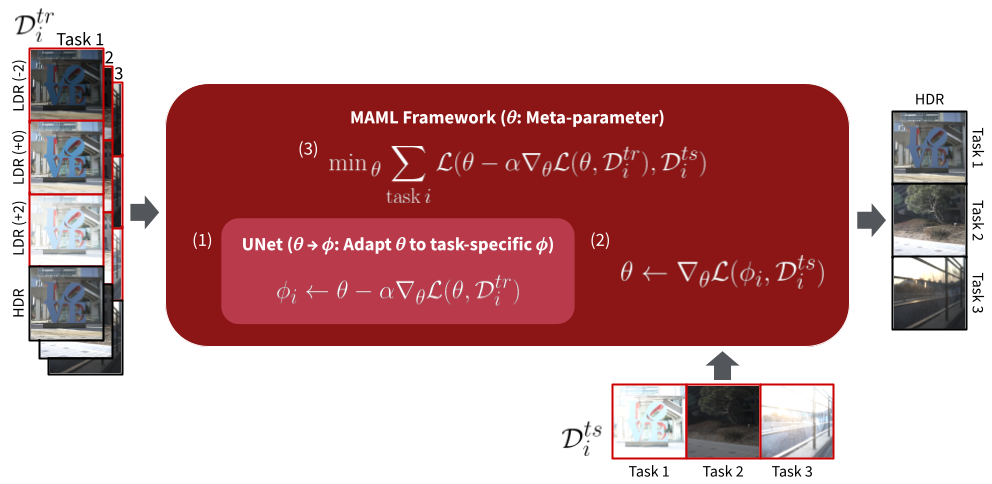}
    \caption{\textbf{MetaHDR architecture}. Given a specific LDR-to-HDR conversion task, MetaHDR uses the training exposures to adapt. Meta parameters $\theta$ are adjusted based on the performance of an unseen test exposure from the same scene.}
    \label{fig:meta-figure-diagram}
\end{figure*}

Later work tries to create an HDR image from a single camera shot. Nayar \textit{et al.} applied an optical encoding mask to the camera sensor in order to capture different exposures of the scene \cite{Nayar}. Then, the recorded pixel values are mapped to an HDR image using a developed image processing algorithm. Additionally, Bist \textit{et al.} applied a variable gamma correction to the image depending on the determined light source in the scene \cite{bist2017tone}.

There have been attempts to pass one LDR image via some deep learning pipeline with the goal of hallucinating an HDR image. In Eilertsen \textit{et al.}, an input LDR image is passed through a deep U-Net style CNN architecture to output an HDR image \cite{Eilertsen_2017}. Santos \textit{et al.} appended an additional fine-tuning stage after passing an LDR image through a deep CNN in order to further correct the over-exposed and under-exposed regions of an image \cite{Marcel:2020:LDRHDR}.

Alternatively, it is possible to combine optical hardware with deep learning to produce an HDR image. In Metzler \textit{et al.}, an optical encoder parameterized by a PSF and a CNN decoder are used to create an end-to-end LDR-to-HDR pipeline that reconstructs HDR images without hallucination \cite{metzler2020deep}. Martel \textit{et al.} developed neural sensors which use neural networks in order to optimize per-pixel shutter functions which can be used to create an HDR image \cite{martel2020neural}.

\section{Methodology}
The overall MetaHDR model is outlined in Fig. \ref{fig:meta-figure-diagram}. Given our task distribution of LDR-to-HDR conversions (i.e. all scenes $\D$), each task $\D_i$ consists of multiple exposures from the same scene as well as corresponding ground truth HDR labels for both task-specific training samples $\Dtr$ and task-specific testing samples $\Dts$. For each iteration, we produce task-specific $\phi_i$ by adapting the meta-parameters $\theta$ using each exposure and HDR pair within $\Dtr$. 

\begin{equation}\label{eq:phi_update}
    \phi_i \leftarrow \theta - \alpha \nabla_\theta \mathcal{L}(\theta, \mathcal{D}^{tr}_i)    
\end{equation}

With our adapted $\phi_i$, we forward pass our test exposure and generate its corresponding HDR counterpart. A loss is computed between the predicted HDR image and the ground truth HDR image. The meta-parameters $\theta$ are then adjusted based on the gradients from the inner update loop. 
\begin{equation}\label{eq:theta_update}
    \theta \leftarrow \nabla_\theta \mathcal{L}(\phi_i, \mathcal{D}_i^{ts})
\end{equation}

Over all iterations of the training procedure, our model weights $\theta$ encapsulate a \textit{general} solution that is close to many nonlinear mappings but specific to none. 
\begin{equation}\label{eq:overall}
    \text{min}_{\,\theta} \sum_{\text{task} \, i} \mathcal{L}(\theta-\alpha\nabla_{\theta}\mathcal{L}(\theta, \mathcal{D}_i^{tr}), \mathcal{D}_i^{ts})
\end{equation}

\subsection{Dataset}
\begin{figure}[htbp!]
    \centering
    \includegraphics[width=0.99\linewidth]{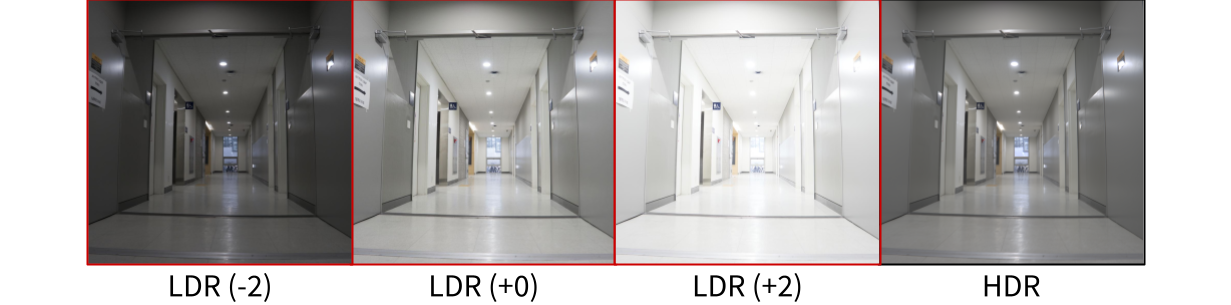}
    \caption{Diagram showing a scene from the dataset. Each scene consists of three LDR images and one HDR image. The red outlines are added for visual clarity.}
    \label{fig:example-input}
\end{figure}

The dataset used \cite{dataset1} had 450 scenes\footnote{At the time of downloading the data, there were 450 scenes. Recently, the author chose to remove some of the images for aesthetic reasons.} where each scene had three LDR images of exposure values -2, 0, +2 and one HDR image. Each image in the dataset was $1024\times1024$. An example set of three LDR images of different exposures and the corresponding HDR image is shown in Fig. \ref{fig:example-input}. 

Because there were only three LDR exposures per scene in the dataset, we attempted to simulate other exposures via linear global tone mapping. We did this by clamping the HDR image to be within a percentile of values, linearly interpolating the pixels within the chosen percentile range to be within 0 and 1, and then applying a gamma correction factor with $\gamma = 2.2$. Although some simulated exposures were promising, sometimes the simulated exposure would saturate too many pixels. This would cause the original HDR image to not be recoverable. Therefore, we decided to use the original dataset without any simulated exposures. Note that due to compute limitations, we cropped each image about the center to $512\times512$.

\subsection{Inner Model Architecture}
Within each task-specific adaptation step, the input images go through a UNet architecture \cite{ronneberger2015u}. This choice was motivated by the performance of existing single-task models, which use UNets to perform HDR reconstruction \cite{Eilertsen_2017}.  

The UNet's contracting layers and expanding layers are organized into \say{blocks} which consist of a sequence of contraction and expansion layers. Each block is a predefined sequence of identical convolutions, activations, maximum-pools, and batch normalizations. The bottom block as well as the top block (i.e. the block that leads to the output image) differ from the other expanding and contracting blocks. At each contracting block, the number of channels is doubled while the number of dimensions is halved. 

The padding, and stride values of each convolution are chosen to facilitate this even dimension reduction. All $3\times3$ convolution operations are done with a stride of 1 and a padding of 1. The universal activation function used in all architecture references to \say{activation} is the ReLU activation function. Max pooling is applied between each contracting block. The following sections detail the specifications of each block in the UNet. 

\paragraph{Contracting Block.}
Each contracting block consists of two 2D $3\times3$ convolution operations terminated with the activation function. The first convolution doubles the number of channels. The second one does not. A 2D batch normalization is done between the convolution operations and the activation function.

\paragraph{Bottom Block.}
The bottom block is an augmented contracting block. After performing the same sequential structure as a contracting block, an extra 2D convolutional upsample is done. This operation halves the number of channels present.

\paragraph{Expanding Block.} 
Taking the concatenated result (which consists of the previous block and a sub-portion of the corresponding contracting block), each expanding block performs one 2D $3\times3$ convolution. This operation halves the number of channels. The second convolution is identical, except it keeps the channel number equal. Finally, the 2D convolutional upsample is performed. Every convolution is followed by the activation function. 

\paragraph{Top Block.}
The top block consists of two convolutions. The first 2D $3\times3$ convolutional layer halves the number of channels. The second does not change the channel number. The final layer is a 2D $1\times1$ convolution that maps the channels to the 3 output RGB channels. The final convolutional layer is followed with a sigmoid activation to force the pixel values to be between 0 and 1. 
        
\subsection{MAML Framework}
MetaHDR uses the learn2learn PyTorch package \cite{L2L}, giving us confidence that the gradient adaptations and updates are implemented correctly. Given that each task consists of two LDR images for adaptation, we characterize the meta-learning problem as a 2-way 1-shot regression problem. We trained MetaHDR with a batch size of 5 tasks, for 200 meta-iterations. Task-specific adaptation was performed over 3 iterations per meta-iterations. The meta-learning learning rate was 0.005 and the task-specific adaptation learning rate was 0.01. 

During meta-test time, adaptation can only occur if users have the ability to generate a plausible set of task-specific adaptation labels. For this purpose, we use an existing single-shot deep CNN based approach to HDR image reconstruction known as HDR-CNN \cite{Eilertsen_2017}. For this reason, the ground truth labels used in task-specific adaptation during meta-training are also generated using the pretrained HDR-CNN model. The impact of training in this manner was studied, and results are discussed in subsequent sections.

\subsection{Loss Function}
The loss function used \cite{expandnet} balanced $\ell_1$ loss with a penalty term for having a low cosine similarity. The $\ell_1$ term forces pixels to be numerically close to each other while the 1 - cosine similarity term forces pixels color similarity between the input and target image.

\begin{equation}\label{eq:loss}
    \mathcal{L}_i = || \hat{I_i} - I_i ||_1 + \lambda \left(1 - \frac{1}{K}\sum_{j=1}^K \frac{\hat{I_i}^j \cdot I_i^j}{||\hat{I_i}^j||_2  ||I_i^j||_2} \right) 
\end{equation}

The loss function, which is referred to as ExpandNet loss \cite{expandnet}, is shown in \ref{eq:loss}. Here, $K$ is the number of pixels in one channel, and ${I_i}^j$ is the rgb-vector at the j-th pixel location for the i-th image and $\hat{I_i}^j$ is the predicted rgb-vector at the j-th pixel location for the i-th image. Note also that the $\ell_1$ term is calculated for every pixel in every channel of the image and then averaged across the entire image. 

We also experimented with using the VGG architecture based Learned Perceptual Image Patch Similarity (LPIPS) metric \cite{lpips} as a loss function. LPIPS is a perceptual similarity metric that uses deep neural networks to capture the many \say{nuances} of human perception.

\section{Results}
We applied our methodology to a held out meta-test set and report our findings here. The meta-test set consisted of 45 scenes of three LDR images and one ground truth HDR image. We evaluated on the model that had the highest meta-validation score during training. For quantitative evaluation, we used SSIM and PSNR values to compare the predicted HDR and the true HDR image. Capturing image similarity within a performance metric is a historically difficult task and is often use-case dependent. Our choice of SSIM and PSNR as performance metrics was based on their general applicability across many different applications.

\subsection{Qualitative Results}

\begin{figure}[htbp!]
     \centering
     \begin{subfigure}{0.95\linewidth}
         \centering
         \includegraphics[width=0.92\linewidth]{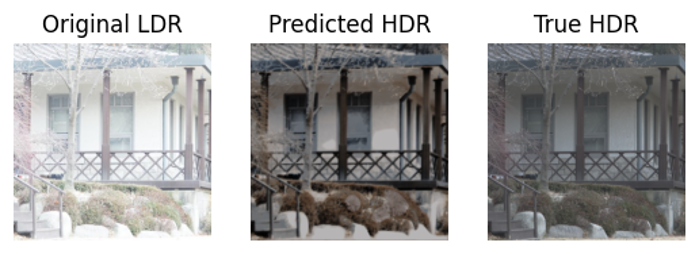}
         \caption{SSIM: 0.686, PSNR: 23.798 dB}
         \label{fig:cabinss}
     \end{subfigure}
     \vfill
     \begin{subfigure}{0.95\linewidth}
         \centering
         \includegraphics[width=0.92\linewidth]{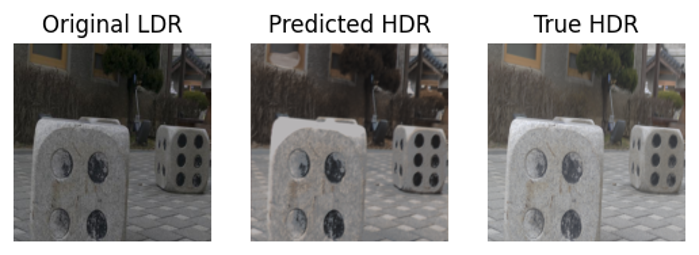}
         \caption{SSIM: 0.648, PSNR: 21.399 dB}
         \label{fig:dicess}
     \end{subfigure}
     \begin{subfigure}{0.95\linewidth}
         \centering
         \includegraphics[width=0.92\linewidth]{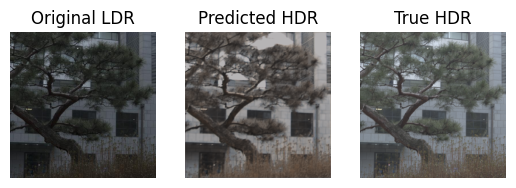}
         \caption{SSIM: 0.788, PSNR: 26.261 dB}
         \label{fig:treess}
     \end{subfigure}
    \caption{Example Single Shot Results}
    \label{fig:single-shot}
\end{figure}

Fig. ~\ref{fig:single-shot} shows example single-shot outputs where we pass each meta-test image through our model without any adaptation at meta-test time.

\begin{figure}[htbp!]
     \centering
     \begin{subfigure}{0.95\linewidth}
         \centering
         \includegraphics[width=0.92\linewidth]{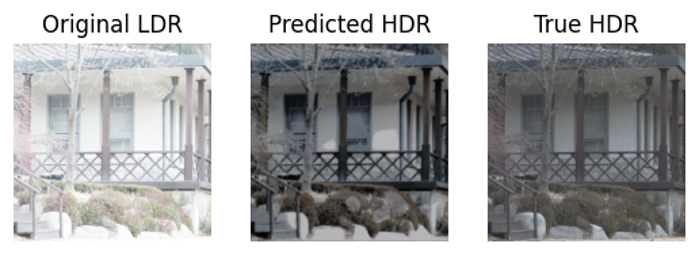}
         \caption{SSIM: 0.703, PSNR: 24.818 dB}
         \label{fig:cabina}
     \end{subfigure}
     \vfill
     \begin{subfigure}{0.95\linewidth}
         \centering
         \includegraphics[width=0.92\linewidth]{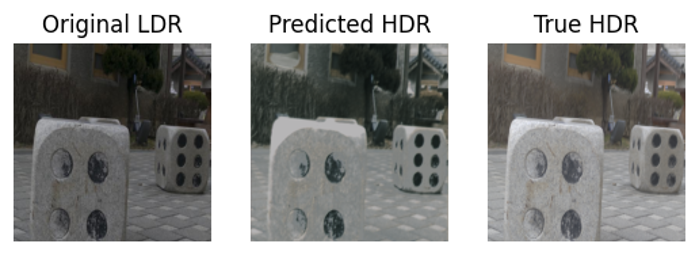}
         \caption{SSIM: 0.768, PSNR: 21.329 dB}
         \label{fig:dicea}
     \end{subfigure}
    \begin{subfigure}{0.95\linewidth}
         \centering
         \includegraphics[width=0.92\linewidth]{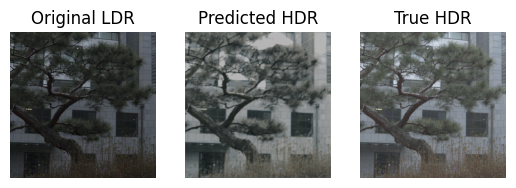}
         \caption{SSIM: 0.791, PSNR: 26.690 dB}
         \label{fig:treea}
     \end{subfigure}
    \caption{Example Adaptation  Results}
    \label{fig:adaptation}
\end{figure}

Fig. ~\ref{fig:adaptation} shows example adaptation outputs. For adaptation at meta-test time, we first simulate what an HDR image may look like using HDR-CNN \cite{Eilertsen_2017}. Then, for two of the three exposures, we adapt the task-specific parameters (from the meta-parameters) for the task at hand using the simulated labels for those two exposures. Finally, for the held out test image, we predict its corresponding HDR image.

In general, we see an improvement from single-shot to adaptation which is expected because the meta-parameters, $\theta$, are specific to no nonlinear mappings and only encapsulate a general solution.

From a visual perspective, the images do fill in many saturated pixels correctly. 
For example, in the cabin in Fig. \ref{fig:cabinss} and Fig. \ref{fig:cabina},  the algorithm was able to fill in the over-exposed regions correctly. Additionally, for the dice in Fig. \ref{fig:dicess} and Fig. \ref{fig:dicea}, the underexposed regions were filled in correctly. However, there are some imperfections. Although the methodology  was able to keep the structure of the tree in Fig. \ref{fig:treess} and Fig. \ref{fig:treea}, it was unable to recover the areas of the window in the background as well as the blue and gray color of the wall. 

\subsection{Quantitative Results}
Evaluation metrics for MetaHDR using different labeling schemes for model adaptation at meta-train and meta-test time can be found in Table \ref{tbl:metaHDR_results}. Additionally, evaluation metrics for MetaHDR using various loss functions can be found in Table \ref{tbl:metaHDR_loss_results}.

\paragraph{Evaluation Results.} 
From Table \ref{tbl:metaHDR_results}, we observe more concretely that adaptation does produce quantitatively better HDR image reconstruction results. Comparing the use of ground truth HDR labels and the use of simulated HDR labels, we see that using simulated HDR labels during meta-training does not significantly hurt quantitative performance (see columns 2,3 and 4,5). Similarly, using simulated HDR labels during meta-test also results in comparable results to using ground truth labels for adaptation (see bottom two rows). These results reinforce the validity of our meta-test HDR label simulation approach. 

\begin{table}[!hbt]\footnotesize 
\begin{center}
\begin{tabular}{|l|*{4}{m{1cm}|}m{1cm}} \hline
 & \multicolumn{2}{c|}{\textbf{SSIM ($\uparrow$)}}  & \multicolumn{2}{c|}{\textbf{PSNR (dB) ($\uparrow$)}} \\ \hhline{>{\arrayrulecolor{white}}-*{4}{>{\arrayrulecolor{black}}-}}
\multirow{-2}{*}{}& Label\cite{debevec_97} & Label\cite{Eilertsen_2017} & Label\cite{debevec_97} & Label\cite{Eilertsen_2017} \\ \hline
LDR No Recon. & \multicolumn{2}{c|}{0.489}  & \multicolumn{2}{c|}{12.200}\\ \hline
 &  & & & \\
\multirow{-2}{*}{Single Shot} & \multirow{-2}{*}{0.666}& \multirow{-2}{*}{0.687}& \multirow{-2}{*}{19.577}& \multirow{-2}{*}{19.713}\\ \hline
 &  & & & \\
\multirow{-2}{*}{\shortstack[l]{Adaptation with \\ True HDR}} & \multirow{-2}{*}{0.683}& \multirow{-2}{*}{0.701}& \multirow{-2}{*}{19.967}& \multirow{-2}{*}{20.125}\\ \hline
 &  & & & \\
\multirow{-2}{*}{\shortstack[l]{Adaptation with \\ Simulated HDR}} & \multirow{-2}{*}{0.687}& \multirow{-2}{*}{0.698}& \multirow{-2}{*}{19.753}& \multirow{-2}{*}{19.790}\\ \hline
\end{tabular}
\caption{Evaluation SSIM and PSNR for scene images reconstructed from the meta-test split of the dataset using various adaptation labeling schemes. Column labels denote the type of HDR label used during meta-training.}
\label{tbl:metaHDR_results}
\end{center}
\end{table}

\paragraph{Loss Function Results.}
At the center of LDR-to-HDR image reconstruction is the question of how to quantitatively measure the similarity between two images. When defining our model, we performed multiple experiments to determine the best loss function to use from a short list of possible contenders (i.e. ExpandNet Loss, LPIPS, and $\ell_1$-regularized LPIPS). Our results show that measuring image similarity using LPIPS as a loss function makes it difficult for the meta-learner to hone in on the common structure in LDR-to-HDR conversion (see Table \ref{tbl:metaHDR_loss_results}). When using LPIPS out of the box, the meta-learner's output after 200 meta-iterations reduces the overall quality of the generated HDR images. We also found that regularizing the LPIPS loss with $\ell_1$ loss led to some improvement relative to its vanilla counterpart. Interestingly, this improvement was only present when using the ground truth labels. Compared to the baseline LDR image metrics, both LPIPS loss based models fail to produce adequate quantitative results. We believe that the cause of this is the combination of SSIM as our performance metric and the use of LPIPS as a loss function. Namely, one of the arguments for using LPIPS is that it better captures structures important for human perception; something that SSIM does not attempt to do.

\begin{table}[!hbt]\footnotesize 
\begin{center}
\begin{tabular}{|l|*{4}{m{1cm}|}m{1cm}} \hline
 & \multicolumn{2}{c|}{\textbf{SSIM ($\uparrow$)}}  & \multicolumn{2}{c|}{\textbf{PSNR (dB) ($\uparrow$)}} \\ \hhline{>{\arrayrulecolor{white}}-*{4}{>{\arrayrulecolor{black}}-}}
\multirow{-2}{*}{}& Label\cite{debevec_97} & Label\cite{Eilertsen_2017} & Label\cite{debevec_97} & Label\cite{Eilertsen_2017} \\ \hline
LDR No Recon. & \multicolumn{2}{c|}{0.489}  & \multicolumn{2}{c|}{12.200}\\ \hline
 &  & & & \\
\multirow{-2}{*}{ExpandNet Loss} & \multirow{-2}{*}{0.687}& \multirow{-2}{*}{0.698}& \multirow{-2}{*}{19.753}& \multirow{-2}{*}{19.790}\\ \hline
 &  & & & \\
\multirow{-2}{*}{LPIPS} & \multirow{-2}{*}{0.362}& \multirow{-2}{*}{0.262}& \multirow{-2}{*}{8.006}& \multirow{-2}{*}{9.226}\\ \hline
 &  & & & \\
\multirow{-2}{*}{LPIPS + $\ell_1$} & \multirow{-2}{*}{0.491}& \multirow{-2}{*}{0.200}& \multirow{-2}{*}{16.122}& \multirow{-2}{*}{8.869}\\ \hline
\end{tabular}
\caption{Evaluation SSIM and PSNR for scene images reconstructed from the meta-test split of the dataset using various loss functions. Results shown after meta-test adaptation, and use HDR-CNN generated meta-test adaptation labels.}
\label{tbl:metaHDR_loss_results}
\end{center}
\end{table}

\paragraph{Comparison to state-of-the-art.}
Existing state-of-the-art single-shot deep learning approaches to HDR image reconstruction generally perform well both quantitatively and qualitatively. As a comparison, we evaluated the HDR-CNN pretrained model \cite{Eilertsen_2017} on the same hidden test set and got a SSIM of 0.704 with PSNR of 18.072 dB. We found this to be comparable to the results we achieved with MetaHDR. One concern we have is the size and scope of the dataset we used. Namely, its small size and pre-partitioned labels were optimal for our purposes. However, there are many possible scenes that have not been represented. Still, we find it encouraging that our model, trained on this small dataset, is able to perform comparably to HDR-CNN, which was trained on a significantly larger and more diverse dataset. 

\section{Discussion}
\subsection{Takeaways} 
In conclusion, our novel approach to data-driven HDR image reconstruction is capable of generating any nonlinear mapping specific to the scene in question. It is able to perform single-shot and few-shot HDR image reconstruction using conventional imaging equipment and achieve performance comparable to state-of-the-art deep learning pipelines for the images tested on. 

The strong baseline single-shot performance of MetaHDR as well as its adaptability when given more exposures from a scene make its utilization uniquely intuitive. One would naturally expect that a model both perform well and perform better when given more data. Indeed, it is neither limited by the need for a wide range of exposures like existing heuristic driven methods \cite{debevec_97} nor limited to one single nonlinear mapping like existing data driven methods\cite{Eilertsen_2017, Marcel:2020:LDRHDR}.

\subsection{Limitations and Challenges}
\begin{figure}[htbp!]
    \centering
    \includegraphics[width=0.95\linewidth]{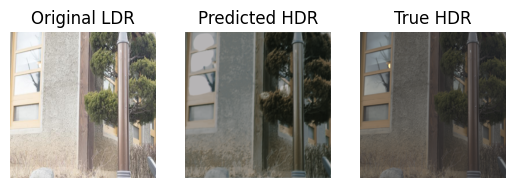}
    \caption{Example of artifacts in reconstructed image. The artifacts are visible in the window of the predicted HDR image.}
    \label{fig:artifact-image}
\end{figure}

Challenges with our model include color retention from input to output and the appearance of unwanted artifacts in the predicted HDR image. As shown in Fig. \ref{fig:artifact-image}, our model was able to successfully recover the image. However, our model added artifacts in the windows which were not there originally. Interestingly, despite these artifacts, the example shown achieved an SSIM of 0.768 and a PSNR of 21.329 dB. This leads us to believe that these issues are more indicative of a need for better performance metrics / loss functions and less indicative of an issue with our meta-learning approach. Adjusting our model to remove artifacts and increase color retention will be especially challenging because these quantitative metrics still believe that the model is performing well.

\subsection{Future Work}
In the future, we would like to explore more loss functions that prioritize color saturation from input to output. This would help remove any artifacts from predicted HDR images as well as improve overall quality of the images. Additionally, we have seen other proposed loss functions specifically tailored for LDR-to-HDR conversion. For example, IRLoss \cite{Eilertsen_2017} uses a weight matrix to force the model to adjust saturated pixels more significantly that non-saturated pixels. This could prove beneficial in giving the model a useful prior to build off of, and alleviate some of our model's challenges.

Furthermore, we have already seen that there is a disconnect between what quantitatively constitutes a \say{good} reconstruction and what qualitatively presents itself as a good reconstruction. One avenue to explore is the use of better performance metrics (like LPIPS) which would ideally help us capture more important image similarity within the loss. 

Finally, we were limited to using 450 scenes with three exposures per scene and one ground truth HDR image. We would like to find or collect a dataset with more scenes and/or more exposures in order to further improve our model. 

\section{Acknowledgments}
This work was completed as a Final Project for EE 367 / CS 448I: Computational Imaging and Display at Stanford University. We would like to thank our professor, Dr. Gordon Wetzstein, for his valuable instruction throughout the quarter and our project mentor, Cindy Nguyen, for giving us useful insights.

{\small
\bibliographystyle{ieee}
\bibliography{metaHDR}
}

\end{document}